\documentclass[twocolumn,prb,aps]{revtex4}
\usepackage{amsmath}
\usepackage{amssymb}
\usepackage{graphicx}
\usepackage{dcolumn}
\usepackage{graphicx}
\usepackage{dcolumn}
\usepackage{bm}

\begin{document}

\preprint{Phys.Rev.B}

\title{Linear magnetoresistance in HgTe quantum wells}
\author{G.M.Gusev,$^1$ E.B Olshanetsky,$^2$  Z.D.Kvon,$^{2,3}$
N.N.Mikhailov,$^2$ and  S.A.Dvoretsky,$^{2}$}

\affiliation{$^1$Instituto de F\'{\i}sica da Universidade de S\~ao
Paulo, 135960-170, S\~ao Paulo, SP, Brazil}
\affiliation{$^2$Institute of Semiconductor Physics, Novosibirsk
630090, Russia}
\affiliation{$^3$Novosibirsk State University,
Novosibirsk, 630090, Russia}

\date{\today}
\begin{abstract}
We report magnetotransport measurements in a HgTe  quantum well with
an inverted band structure, which  is expected to be a
two-dimensional (2D) topological insulator. A small magnetic field
perpendicular the 2D layer breaks the time reversal symmetry and
thereby, suppresses the edge state transport. A linear
magnetoresistance is observed in low magnetic fields, when the
chemical potential moves through the the bulk gap. That
magnetoresistance is well described by numerical calculations of the
edge states magnetotransport in the presence of nonmagnetic
disorder. With magnetic field increasing the resistance, measured
both in the local and nonlocal configurations first sharply
decreases and then increases again in disagreement with the existing
theories.


\end{abstract}

\maketitle

The topological insulators are a novel type of system with a gap
in the energy spectrum of the bulk states and a gapless energy
spectrum of a special class of electron states located at their
surface or edges \cite{hasan, hasan2, qi}. The two-dimensional
(2D) topological insulator (TI) has gapless states propagating
along its edges\cite{konig, maciejko1}. There are two famous
examples of such system: the quantum Hall effect (QHE) state,
which exists in a strong magnetic field perpendicular to the plane
and is characterized by chiral edge states, and the time-
reversal-symmetric quantum spin Hall effect state (QSHE), which is
induced by a strong spin-orbit (SO) interaction and is
characterized by counter propagating states with opposite spins in
the absence of magnetic field.

The QSHE has been realized in HgTe/CdTe quantum wells with
inverted band structure \cite{konig2, buhmann}. The existence of
edge channel transport in the QSH regime has been proved
experimentally \cite{konig}, when a 4-probe resistance in an
HgTe/CdTe micrometer-sized ballistic Hall bar demonstrated a
quantized plateaux $R_{xx}\simeq h/2e^{2}$. One more experimental
evidence for edge states in QSHE is a nonlocal transport, since
the application of the current between any pair of the probes
creates net current along the sample edge, and can be detected by
any other pair of the voltage probes. \cite{roth, gusev}. It is
expected that the stability of helical edge states in the
topological insulator is unaffected by the presence of weak
disorder \cite{hasan, qi, moore, moore2, kane, bernevig,
bernevig2}. Note however, that the quantized ballistic transport
has been observed only in micrometer-sized samples, and plateaux
$R_{xx}\simeq h/2e^{2}$ is destroyed, if the sample siza is above
a certain critical value about a few microns \cite{konig}.
\begin{figure}[ht!]
\includegraphics[width=7cm,clip=]{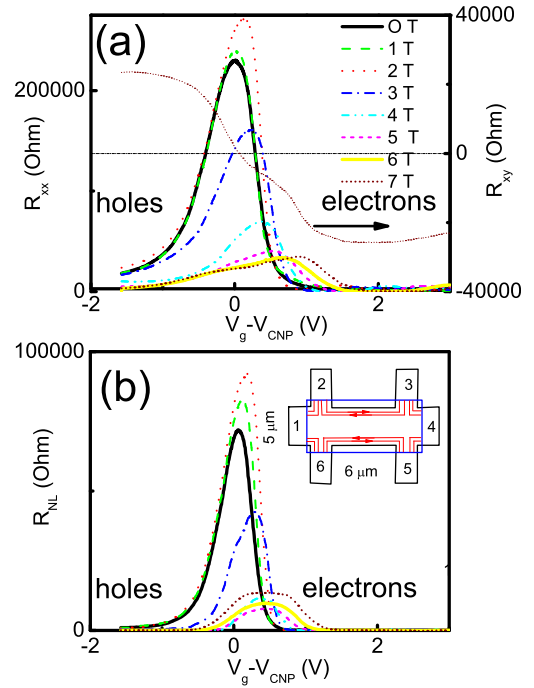}
\caption{\label{fig.1}(Color online) (a) The longitudinal $R_{xx}$ (
I=1,4; V=2,3) and Hall $R_{xy}$ ( I=1,4; V=3,5) resistances as a
function of the gate voltage at zero and different nonzero magnetic
fields, T=4.2 K. (b)The nonlocal $R_{NL}$ ( I=1,2; V=3,5) resistance
as a function of the gate voltage at zero and different nonzero
magnetic fields. The inset -(b) Top view of the sample. The gate is
shown by rectangle.  }
\end{figure}

A magnetic field perpendicular to the 2D layer breaks time
reversal symmetry (TRS) and thereby enables elastic scattering
between counterpropagating chiral edge states. However, a number
of the different theoretical models has previously been proposed
\cite{konig,tkachov,scharf,chen,maciejko2} and different
conflicting scenarios have been developed for TRS breaking in QHSE
system, which requires detailed experimental investigation.

A sharp magnetorsistance spike has been observed in the previous
study of a HgTe-based sample few microns in size \cite{konig}.
Nevertheless, it is very likely that these experiments have been
done in the regime, when the disorder strength $W$ is of the same
order or even larger than the bulk energy gap $E_{g}$. In this
paper we report on the observation and a systematic investigation
of a positive linear magnetoresistance in HgTe quantum wells with
inverted band structure corresponding to the QSHE phase. The
magnetoresistance in low fields is described by a theoretical
model \cite{maciejko2} that takes into account the combined effect
of disorder and TRS breaking in a weak disorder regime, when
$W<E_{g}$. In magnetic fields above 2 T we observe a decrease of
the resistance with a saturation, corresponding to the QHE phase,
followed by a transition to a state with a higher resistance and
nonlocal transport in fields above 6 T.

The $Cd_{0.65}Hg_{0.35}Te/HgTe/Cd_{0.65}Hg_{0.35}Te$ quantum wells
with the (013) surface orientations and the width $d$ of 8-8.3 nm
were prepared by molecular beam epitaxy. A detailed description of
the sample structure has been given in \cite{kvon,gusev2,
olshanetsky}. The six-probe Hall bar was fabricated with the
lithographic length $6 \mu m$ and the width $5 \mu m$ (Figure 1,
insert). The ohmic contacts to the two-dimensional gas were formed
by the in-burning of indium. To prepare the gate, a dielectric
layer containing 100 nm $SiO_{2}$ and 200 nm $Si_{3}Ni_{4}$ was
first grown on the structure using the plasmochemical method.
Then, a TiAu gate with the size $18\times10 \mu m^{2}$ was
deposited. Several devices with the same configuration have been
studied. The density variation with gate voltage was $1.09\times
10^{15} m^{-2}V^{-1}$. The magnetotransport measurements in the
described structures were performed in the temperature range
1.4-25 K and in magnetic fields up to 12 T using a standard four
point circuit with a 3-13 Hz ac current of 0.1-10 nA through the
sample, which is sufficiently low to avoid the overheating
effects.

The carriers density in HgTe quantum wells can be varied
electrostatically with the gate voltage $V_{g}$. The typical
dependence of the four-terminal $R_{xx}=R_{I=1,4;V=2,3}$ and Hall
$R_{xy}=R_{I=1,4;V=3,5}$ resistances of one representative sample
as a function of $V_{g}$ is shown in Figure 1a. The resistance
$R_{xx}$ exhibits a sharp peak that is $\sim 20$ times greater
than the universal value $h/2e^{2}$, which is expected for QSHE
phase. This value varies from 150 to 300 kOhm in different
samples.  The Hall coefficient reverses its sign and
$R_{xy}\approx0$ when $R_{xx}$ approaches its maximum value
\cite{gusev}, which can be identified as the charge neutrality
point (CNP). This behavior resembles the ambipolar field effect
observed in graphene \cite{sarma}. The gate voltage induces charge
density variations, transforming the quantum well conductivity
from n-type to  p-type via a QHSE state.

As we mentioned above, that the edge state transport is unaffected
by the presence of weak disorder \cite{hasan, qi, moore,
moore2,kane,bernevig, bernevig2}. However, the quantized ballistic
transport  and plateaux $R_{xx}\simeq h/2e^{2}$ have not been
observed in samples with dimensions above a few microns
\cite{konig}. One of possible explanations is presence of local
fluctuations of the energy gap induced by smooth inhomogeneities,
which can be represented as metallic islands. In accordance with
Landauer-B$\ddot{u}$ttiker formalism \cite{buttiker} any voltage
probe coupled to a coherent conductor introduces incoherent
inelastic processes and modifies the ballistic transport. Metallic
islands can result in similar effects, since an electron entering
the island is dissipated and thermalized there and later on fed back
into the system. Therefore ballistic coherent transport is expected
only in the region between the islands, and the total 4-terminal
resistance exceeds the quantized value. However, such long range
potential fluctuations must have the amplitude of the order of the
energy gap $E_{g}\sim 30 meV$ which is very unlikely, since such
fluctuations should suppress the electron SdH oscillations in low
magnetic fields, which is not observed in the experiment. the
resistance of samples longer than $1 \mu$m  might be  much higher
than $h/2e^2$  due to  the presence of the spin dephasing (electron
spin flip backscattering on each boundary)\cite{jiang}. Mechanisms
of the back scattering are new and appealing task for theoreticians
and is a matter of ongoing debate. The classical and quantum
magnetic impurities may introduce a backscattering between counter
propagating channels. An accidently formed quantum dot with an odd
number of trapped electrons could play a role of such magnetic
impurity. For strong enough electron-electron interaction the
formation of Luttinger liquid insulator with a thermally activated
transport was predicted \cite{maciejko3}. In the frames of a
somewhat different approach an edge state transport theory in the
presence of spin orbit Rashba coupling has been developed
\cite{strom}. According to this model the combination of a spatially
non-uniform Rashba spin-orbit interaction and a  strong
electron-electron interaction leads to localization of the edge
electrons at low temperatures. However, the exact examination and
comparison with theoretical models requires a further experimental
investigation of the temperature, doping and disorder dependence of
the resistivity which are out of the scope of the present paper.

\begin{figure}[ht!]
\includegraphics[width=7cm,clip=]{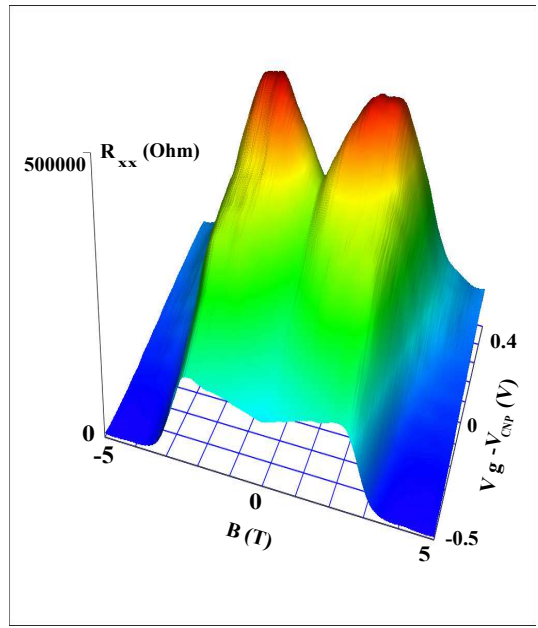}
\caption{\label{fig.2}(Color online)
 The longitudinal resistance $R_{xx}$  as a function
of the gate voltage and magnetic field, T=1.4 K. }
\end{figure}

Figure 1 b shows the nonlocal resistance $R_{NL}$ corresponding to
the configuration, when the current flows between contacts 1 and 2
 and the voltage is measured between contacts 3 and 5. One can see that the
nonlocal resistance $R_{NL}=R_{I=1,2;V=3,5}$ in the topological
insulator phase has a peak of a comparable amplitude, though less
wide, and approximately in the same position as the local
resistance. Outside of the peak the nonlocal resistance is
negligible. One can see that the evolution of resistances with
magnetic field is practically the same in both cases: resistance
grows with field below 2 T, and then rapidly decreases and
saturates. Figure 2 shows the longitudinal resistance $R_{xx}$ in
the voltage-magnetic field plane. One can see the evolution of the
resistance $R_{xx}$ with magnetic field and density, when the
chemical potential crosses the bulk gap. The magnetorsistance
demonstrates a striking V-shape dependence in magnetic fields
below 1T. It is worth noting that the V-shaped magnetoresistance
is observed almost anywhere on the hole side of the peak and
rapidly disappears on the electronic side.

In magnetic fields above 2 T the magnetoresistance starts to
decrease marking a pronounced crossover to the quantum Hall effect
regime. Note, however, that the resistance does not turn to zero,
as would be expected for a conventional QHE state, but approaches
the value $R_{xx}\approx h/e^{2}$. Figure 3 shows the magnetic
field dependence extended to 12 T of the local and nonlocal
resistances at the gate voltage corresponding the peak maximum for
another representative sample. Both the local and nonlocal
resistances grow rapidly in fields above 6 T.  The evolution of
the magnetoresistance in strong quantized magnetic field disagrees
with the theoretical models proposed recently for transport in
HgTe quantum wells with inverted band structure \cite{scharf,
chen}. The growth of the local and nonlocal resistances in the
field above 6 T  can be attributed to the edge state transport via
counter propagating chiral modes similar to the HgTe semimetal
\cite{gusev3} and graphene \cite{abanin} near $\nu=0$. Further
theoretical work would be needed in order to explain this
behaviour.

\begin{figure}[ht!]
\includegraphics[width=7cm,clip=]{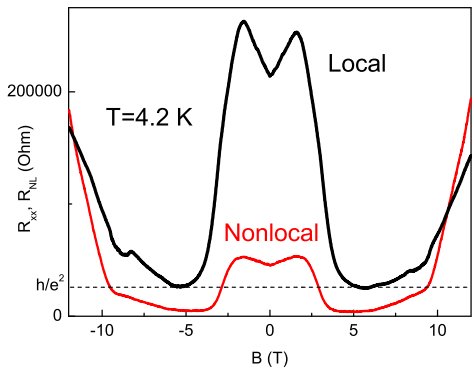}
\caption{\label{fig.3}(Color online) The local $R_{xx}$ (a)  and
nonlocal $R_{NL}=R_{I=2,6;V=3,5}$  resistances as a function of the
magnetic field near the peak maximum (CNP), T=4.2 K.}
\end{figure}

Figure 4 shows the low field part of the relative
magnetoconductivity $\sigma_{xx}(B)/\sigma_{xx}(0)$ for two values
of the gate voltage, one at the CNP and  the other just slightly
below the CNP on the electron side of the TI peak. The
conductivities have been recalculated from experimentally measured
$\rho_{xx}$ and $\rho_{xy}$ by tensor inversion. Note, however, that
around the CNP $\rho_{xy}<<\rho_{xx}$ and
$\sigma_{xx}\approx1/\rho_{xx}$.

\begin{figure}[ht!]
\includegraphics[width=6cm,clip=]{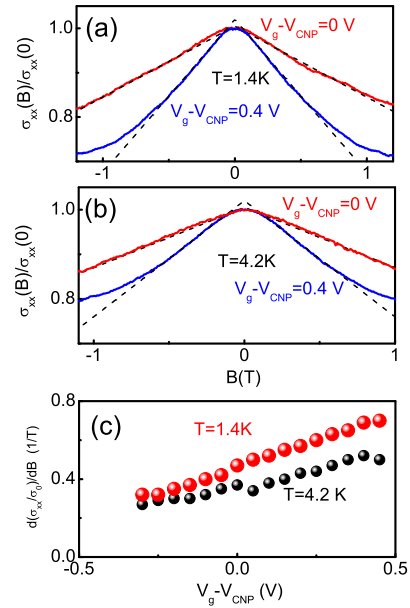}
\caption{\label{fig.4}(Color online) The relative
magnetoconductivity for two values of gate voltage $V_{g}$ and two
temperaures T=1.4 K (a) and T=4.2 K(b). Dashed-lines are the
functions $\sigma_{xx}(B)/\sigma_{xx}(0)=-\alpha|B|$. (c) The
parameter $\alpha\approx d[\sigma_{xx}(B)/\sigma_{xx}(0)]dB$ as a
function of the gate voltage for two temperatures: T=1.4 K (large
circles) and T=4.2 K ( small circles). }
\end{figure}

One can see that the low field part of the conductivity is
describe by the linear function
$\sigma_{xx}(B)/\sigma_{xx}(0)=-\alpha|B|$, where parameter
$\alpha\approx d[\sigma_{xx}(B)/\sigma_{xx}(0)]dB$ is slightly
dependent on temperature and gate voltage (Figure 4c). It is worth
noting that in a small region around zero the magnetoconductance
shows a parabolic behaviour.

In the rest of the paper we will focus on the explanation of the
cusp-like feature in the magnetoconductance near the CNP. As
mentioned in the introduction, the gapless  edge states are
protected from scattering by TRS, which results in a robust
ballistic transport. The 4 terminal resistance in our samples with
the gate size $18\times10 \mu m^{2}$ $R_{xx}\simeq 300kOhm$ is still
significantly higher than $h/2e^{2}$. We have already mentioned
above that this discrepancy is not understood. A finite magnetic
field breaks down the TRS, and the transport
 of the edge states is strongly suppressed. However, different models predict
 substantially different physical scenarios.

For example, one of the model predicts that the external magnetic
field opens a gap in the edge states dispersion \cite{konig}. The
gap is rather small ($E_{g}\sim 0.3 meV$ at B=0.1 T) in accordance
with theoretical estimations, and conductance should be suppressed
in a very narrow interval of the energy, when the chemical potential
goes through this gap. Experimentally, though, a suppression of
conductance is observed in a much wider interval of the carrier
densities, corresponding to the passage of the chemical potential
through the bulk gap $E_{g}\sim 30 meV$, in contrast to the
theoretical predictions. In \cite{tkachov} it is predicted that the
counter-propagating helical edge states persist in a strong magnetic
field. In this model the magnetic field does not create a gap.
Instead, it modifies the energy spectrum of the edge states: one of
the state merges with the lower bulk Landau level, while the other
one remains unchanged. This transformation generates backscattering
between the counter propagating modes in the presence of the weak
disorder and leads to an increase in the resistance. However, the
model \cite{tkachov} does not suggest any realistic description of
the scattering and can hardly be compared with experimental
observations.

 The third model
\cite{scharf} also claims that edge states persist in relatively
low magnetic fields, but in magnetic fields above a certain
critical ($B_{c}$), the band structure becomes normal and the
system turns into an ordinary insulator. For HgTe devices the
critical magnetic field is estimated as $B_{c}\approx7 .4 T$,
therefore the resistance increase at fields above 7 T (figure 3)
can be attributed to the TI- ordinary insulator transition.
However, the model can not explain the growth of the nonlocal
resistance at high magnetic field. A similar, but a somewhat more
complicated evolution of the energy spectrum with B has been
proposed in \cite{chen}.

Finally a numerical study of the edge state transport  in the
presence of both disorder and magnetic field has been reported
recently in Ref.18. The authors predict a negative linear
magnetoconductance $\frac{\Delta G}{e^{2}/h}=-A|B|$, where
parameter A strongly depends on the disorder strength W. A
physical interpretation of the linear magnetoreconductance  is
given along with the effects analogous to the one dimensional (1D)
or 2D antilocalizaton. We believe that the theoretical model
\cite{maciejko2} describing the effect of the disorder and TRS
breaking on the edge transport correctly explains the linear
magnetoresistance observed in our experiment. The theory considers
two regimes: one corresponding to a weak disorder, when $W<E_{g}$,
and the edge states are described by spinless 1D edge liquid, and
a strong disorder regime, when $W>E_{g}$ and the edge states can
penetrate deeper into the bulk. Sensitivity to magnetic field
strongly depends on which of the two regimes is realized:
parameter $A$ is small for weak disorder and abruptly increases by
almost 10-100 times for $W>E_{g}$. Supposing that the results are
valid for a nonballistic case and $A\sim \alpha$, we may conclude
that in our samples $W<E_{g}$. Unfortunately the precision of the
numerical calculations in \cite{maciejko2} does not allow an
unambiguous determination of the disorder parameter $W$ from the
B-slope of the magnetoconductance. It is worth noting that the
B-slope of the sharp magnetoconductance spike observed in samples
with similar size in \cite{konig} is 120 times larger than that in
our samples. Admitting that the model \cite{maciejko2} is
applicable to this data, we obtain the disorder parameter
$W\approx 72 meV$, which is almost 2 time larger than the energy
gap $E_{g}=40 meV$. The disorder parameter $W$ is related to the
local deviations of the HgTe quantum well thickness from its
average value \cite{tkachov2}, rather than to the random potential
due to charged impurities. As has been shown in \cite{maciejko2},
parameter $W$ can be estimated from the value of the mobility. For
example, the mobility $\mu \approx 10^{5} cm^{2}/Vs$ corresponds
to the momentum relaxation time $\tau=0.57 ps$, which can be
derived from the equation $\tau=\hbar/2 \pi \nu (Wa)^{2}$, where
$a=30 \AA$ is the range of the disorder, $\nu=m^{*}/\pi
\hbar^{2}$, $m^{*}$ is the effective mass. Substituting these
parameters into the equation for the relaxation time yields $W=22
meV$. It is worth noting that the average band gap can be smaller
due to the stress. For example the energy gaps considered in
\cite{tkachov2} were $E_{g}=14 meV$ for the well width $d=7.3 nm$
and $E_{g}\approx 20 meV$ for $d=8 nm$. This may explain the
difference between our results and those obtained previously in
narrow samples \cite{konig}: the fluctuations $W\sim 15 meV$
result in a large B-slope in wells with $d=7.3 nm$ corresponding
to a strong disorder regime in these wells ($W>E_{g}$) and to a
small B-slope for wider a well $d=8 nm$ ($W<E_{g}$).

The physcial mechanism behind the linear magnetoconductance can
not be unambiguously identified from the numerical calculations
\cite{maciejko2}. It is expected, that this mechanism is rather
related to a suppression of the interference between closed paths,
than the orbital effect, and is analogous to the 1D or 2D
antilocalization. Weak temperature dependence observed in our
experiments supports this interpretation. The authors claim that
for weak disorder the magnetic field has only a perturbative
effect on the transport properties of the edge states and expect a
quadratic dependence of the magnetoconductane on B, rather than
linear. However, it is not evident from the figures, as has been
mentioned by the authors themselves. Further theoretical study
will be needed to better understand the mechanism of TRS breaking
and the effect of disorder on the edge transport in the QSHE.

In conclusion, we have observed a linear negative
magnetoconductance in HgTe-based quantum wells in the QSHE regime,
when the edge state transport prevails. Our observation agrees
with the numerical calculations of the magnetoconductance due to
the edge states transport in the presence of nonmagnetic disorder.
The B-slope of the magnetoconductance is small and corresponds to
a weak disorder limit, when $W<E_{g}$ and the magnetoconductance
is analogous to a one dimensional antilocalizaton. In magnetic
field above 2 T the resistance rapidly decreases and then
saturates, which corresponds to the QHE phase. Above 6 T a
transition to high resistance state is observed, accompanied by a
large nonlocal response, which disagrees with the theory.

 We thank O.E.Raichev for helpful discussions. A financial support of
this work by FAPESP, CNPq (Brazilian agencies), RFBI (N 12-02-00054
and N 11-02-12142-ofi-m) and RAS programs "Fundamental researches in
nanotechnology and nanomaterials" and "Condensed matter quantum
physics" is acknowledged.

\end{document}